\documentclass[twocolumn,prb,showpacs,superscriptaddress,floatfix,nofootinbib]{revtex4}
\usepackage{amssymb}
\usepackage{amsmath}
\usepackage{graphicx}
\usepackage{dcolumn}
\usepackage{bm}
\usepackage[usenames]{color}
\usepackage{ulem}

\DeclareMathOperator{\sign}{\mathrm{sign}}
\DeclareMathOperator{\Real}{\mathrm{Re}}
\DeclareMathOperator{\Imag}{\mathrm{Im}}

\newcommand{\units}[1]{\ensuremath{\,\mathrm{#1}}}

\newif\ifcom
\newif\ifdel
\comtrue
\deltrue
\begin{document}

\title{Superconductor-insulator-ferromagnet-superconductor Josephson
     junction: From the dirty to the clean limit}
\author{N.~G.~Pugach}
\email{pugach@magn.ru}
\affiliation{Faculty of Physics, M.V. Lomonosov Moscow State University, 119992 Leninskie
Gory, Moscow, Russia}
\author{M.~Yu.~Kupriyanov}
\affiliation{Skobeltsyn Nuclear Physics Institute, M.V. Lomonosov Moscow State
University, 119992 Leninskie Gory, Moscow, Russia}
\author{E.~Goldobin}
\author{R.~Kleiner}
\author{D.~Koelle}
\affiliation{Physikalisches Institut--Experimentalphysik  II  and Center for Collective
Quantum Phenomena, Universit\"at T\"ubingen, Auf der Morgenstelle 14,
D-72076 T\"ubingen, Germany }
\date{\today }

\begin{abstract}
The proximity effect and the Josephson current in a
superconductor-insulator-ferromagnet-superconductor (SIFS) junction are
investigated within the framework of the quasiclassical Eilenberger
equations. This investigation allows us to compare the dirty and the clean
limits, to investigate an arbitrary impurity scattering, and to determine
the applicability limits of the Usadel equations for such structures. The
role of different types of the FS interface is analyzed. It is shown that
the decay length $\xi_1$ and the spatial oscillation period $2\pi\xi_2$ of
the Eilenberger function may exhibit a nonmonotonic dependence on the
properties of the ferromagnetic layer such as exchange field or electron
mean free path. The results of our calculations are applied to the
interpretation of experimentally observed dependencies of the critical
current density on the ferromagnet thickness in Josephson junctions
containing a Ni layer with an arbitrary scattering.
\end{abstract}

\maketitle

\section{Introduction}

\label{sec:Int}

Superconductor-ferromagnet-superconductor (SFS) Josephson junctions are a
subject of intensive theoretical and experimental studies \cite%
{KuprReview,Buzdin:2005:Review:SF,VolkovRevModPhys}. In particular, the
question of the applicability range of predictions from dirty and clean
limit theories and the treatment of the crossover between these two limits
has been recognized as an important problem for the theoretical description
of SFS structures \cite{Volkov,Linder}.

For the majority of experimental realizations of SFS structures \cite%
{Ryazanov:2001:SFS-PiJJ,RyazanovBuzdinJc(d)PRL2006,Blum:2002:IcOscillations,Sellier:2003:SFS,Born:2006:SIFS-Ni3Al,Weides:2006:SIFS-HiJcPiJJ,Weides:0-piLJJ,Pfeiffer2008:SIFS:T-Esc,RobinsonPRL97,RobinsonPRB76}
 the exchange energy $H$ of the ferromagnetic materials is rather large \cite%
{Born:2006:SIFS-Ni3Al, RobinsonPRL97}. As a consequence, the characteristic
magnetic length $\xi _{H}=\hbar v_{f}/2H\lesssim \ell_f$, where $\ell_f$ is
the electron mean free path and $v_{f}$ is the Fermi velocity in the F layer
\cite%
{Ryazanov:2001:SFS-PiJJ,RyazanovBuzdinJc(d)PRL2006,Blum:2002:IcOscillations,Sellier:2003:SFS,Born:2006:SIFS-Ni3Al,Weides:2006:SIFS-HiJcPiJJ,Weides:0-piLJJ,Pfeiffer2008:SIFS:T-Esc,RobinsonPRL97,RobinsonPRB76}%
. Under this condition the numerous theoretical predictions \cite%
{KuprReview,Buzdin:2005:Review:SF,VolkovRevModPhys} based on the Usadel
equations \cite{Usadel} have a rather restricted range of validity; i.e.~a
more general approach based on the Eilenberger equations \cite{Eilenberger}
has to be developed.

A simple expression for the critical current density $J_{C},$ which is valid
in the clean limit, was derived in Ref. \onlinecite{Buzdin:1982}.  Still the
analysis of the Eilenberger equations for more general cases, remained a
difficult problem. A significant progress along this direction has been
achieved a decade ago \cite{Volkov}, where the solution of the Eilenberger
equations for arbitrary scattering has been expressed in an integral form
\cite{Volkov}. It was supposed \cite{Volkov} that the SF interface
transparency $D$ is small enough, providing the opportunity to use the
linearized equations. Thus, a general expression for the Josephson junction
supercurrent has been derived and used for numerical evaluations. The
integral representation of the solution of the Eilenberger equations also
permits to reproduce analytical expressions for $J_{C}$ obtained earlier
within both clean and dirty limits \cite%
{Buzdin:1982,BuzdinKuprJETP1992,Ryazanov:2001:SFS-PiJJ}. Recently, the same
results have been achieved using the Ricatti parametrization of the
one-dimensional Eilenberger equations \cite{Linder}. However, the obtained
expressions \cite{Volkov} are so complicated that they are difficult analyze
and use in practice. Although the use of a one-dimensional equation \cite%
{Linder} significantly simplifies the problem, it contains a
non-controllable assumption about the insignificance of the angular
distribution of the Eilenberger function.

Despite of this progress, it is still not clear within which range of
parameters and with what accuracy it is possible to use simple expressions
for the clean \cite{Buzdin:1982} and dirty \cite{BuzdinKuprJETP1992} limits.
The question about the influence of transport properties of SF interfaces is
also still open. Since the answer to these questions is rather important for
experimentalists, we formulate here a problem for a particular case of
superconductor-insulator-ferromagnet-superconductor (SIFS) junctions.
In comparison with SFS, in SIFS structures superconductivity is induced
in the F layer only from one S electrode. This essentially simplifies
the analysis compared to that for SFS junctions.

On the one hand, SIFS junctions are interesting by themselves as potential
elements for superconducting logic circuits \cite%
{Terzioglu:1998:CJJ-Logic,Ioffe:1999:sds-waveQubit,UstinovKapl2003} because
it is possible to vary their critical current density in the $\pi$ state
within a rather wide range, still keeping a high $J_{C}R_{N}$ product \cite%
{Weides:0-piLJJ,Pepe}, where $R_{N}$ is the junction normal resistance per
square.
Furthermore, in comparison with SFS junctions,
SIFS junctions have very small damping which decreases exponentially
at $T \to 0$. This makes them useful for superconducting circuits where
low damping is required.
On the other hand, an SIFS junction represents a convenient model
system for a comparative study of 0-$\pi$ transitions for an arbitrary ratio
of characteristic lengths in the F layer: the F-layer thickness $d_f$, the
mean free path $\ell_f$, the characteristic magnetic length $\xi_H$, and the
nonmagnetic coherence length $\xi_0=\hbar v_{f}/2\pi k_B T,$ where $T$ is
the temperature.

To calculate $J_{C}$ of an SIFS junction, it is sufficient to study the
proximity effect in the FS bilayer and to calculate the magnitude of the
Eilenberger functions at the IF interface. For simplicity we will restrict
ourselves to the study of the situation when the anomalous Green's function
induced in the F layer is small enough, permitting us to use the linearized
Eilenberger equation for the description of the superconducting properties
induced in the ferromagnet. Such an approximation is valid if the FS
interface has a small transparency or if $T$ is close to the critical
temperature $T_{c}$ of the S electrodes.

This article is organized as follows. In sec.~\ref{Sec:Model} we describe
our model based on the linearized Eilenberger equation supplemented with
Zaitsev boundary conditions \cite{Zaitsev} for an SIFS junction. Different
types of FS boundaries are analyzed. Section \ref{Sec:Discussion} presents
dependencies of anomalous Green's functions and the critical current density
on the F-layer parameters. The comparison with experimental results  is
presented in Section \ref{Sec:Exper}. Section \ref{Sec:Conclusion} concludes
this work. The calculation details can be found in the Appendix.

\section{Model}

\label{Sec:Model}

\subsection{Proximity effect in the FS electrode}

To analyze the proximity effect in an FS bilayer for arbitrary values of the
electron mean free paths $\ell _{s}$ and $\ell _{f}$ in the S and F layer,
respectively, it is convenient to introduce the following functions
\begin{equation}
\Phi _{+}=\frac{1}{2}\left[ f(x,\theta _{s},\omega )+f^{+}(x,\theta
_{s},\omega )\right] ,  \label{Fplus}
\end{equation}%
\begin{equation}
\Phi _{-}=\frac{1}{2}\left[ f(x,\theta _{s},\omega )-f^{+}(x,\theta
_{s},\omega )\right] ,  \label{Fminus}
\end{equation}%
where $f(x,\theta _{s},\omega )$ and $f^{+}(x,\theta _{s},\omega )$ are the
quasiclassical Eilenberger functions \cite{Eilenberger}. Then we rewrite the
Eilenberger equations for the S layer, located at $\ 0\leq x\leq \infty $,
in the form:
\begin{equation}
\frac{k_{s}^{2}}{\ell _{s}^{2}\mu _{s}^{2}}\frac{\partial ^{2}}{\partial
x^{2}}\Phi _{+}-\Phi _{+}=-\frac{2\Delta \tau _{s}+\left\langle \Phi
_{+}\right\rangle }{2\left\vert \omega \right\vert \tau _{s}+1},
\label{FplusinS}
\end{equation}%
where
\begin{equation}
\left\langle \Phi _{+}\right\rangle =\int_{0}^{1}\Phi_{+} d\mu,
\label{Fi_av}
\end{equation}%
\begin{equation}
\Phi _{-}=-\frac{\ell _{s} \sign (\omega )}{k_{s}}\mu _{s}\frac{\partial }{%
\partial x}\Phi _{+}\;.  \label{FminusinS}
\end{equation}%
The insulating layer of negligible thickness is located at $x=-d_{f}$. In
the F layer (located at $-d_{f}\leq x\leq 0$) the equations have the form
\begin{equation}
\frac{k_{f}^{2}}{\ell _{f}^{2}\mu _{f}^{2}}\frac{\partial ^{2}}{\partial
x^{2}}\Phi _{+}-\Phi _{+}=-\frac{\left\langle \Phi _{+}\right\rangle }{k_{f}}%
,  \label{FplusinF}
\end{equation}%
\begin{equation}
\Phi _{-}=-\frac{\ell _{f} \sign (\omega )}{k_{f}}\mu _{f}\frac{\partial }{%
\partial x}\Phi _{+}.  \label{FminusinF}
\end{equation}%
Here $\omega =\pi T(2n+1)$ are Matsubara frequencies, $\theta _{s,f}$ are
the angles between the FS interface normal and the direction of Fermi
velocities $v_{s,f}$ in the S and F layers, respectively, $\mu
_{f,s}=\cos\theta _{f,s},$ $\Delta$ is the superconducting order parameter,
which is assumed to be zero in the ferromagnet, $\tau_{s,f}=\ell
_{s,f}/v_{s,f}$ are electron scattering times, and corresponding wave
vectors $k_{s}=2\left\vert \omega \right\vert \tau _{s}+1,$ $%
k_{f}=2[\left\vert \omega \right\vert +iH \sign(\omega )]\tau
_{f}+1=2\left\vert \omega \right\vert \ell _{f}/v_{f}+i2H\ell _{f}/v_{f} %
\sign(\omega )+1=1+\ell _{f}/\xi _{\omega }+i \sign(\omega )\ell _{f}/\xi
_{H},$ $\xi _{\omega }=v_{f}/2\left\vert \omega \right\vert .$ We use the
units where $\hbar=1$ and $k_B=1$. Here we also take into account that the
normal Eilenberger function in the ferromagnet $g\approx \sign(\omega ).$ We
also neglect multiple reflections from the FS and the IF interfaces, which
is reasonable if $d_{f} \geqslant \ell _{f}$, i.e. the ballistic regime is
not considered. The fully ballistic case with a double barrier SIFIS
junction was examined in Ref.~[\onlinecite{Radovic2003}].

Equations (\ref{FplusinS})--(\ref{FminusinF}) must be supplemented by the
boundary conditions. At $x\rightarrow \infty $ the function $\Phi _{+}$
should approach its bulk value
\begin{equation}
\Phi _{+}(\theta _{s},\omega )=\frac{\Delta _{0}}{\sqrt{\Delta
_{0}^{2}+\omega ^{2}}},\   \label{Bcds}
\end{equation}%
where $\Delta _{0}$ is the magnitude of superconductor order parameter far
from the SF\ interface. At $x=-d_{f}$ the boundary condition
\begin{equation}
\frac{d}{dx}\Phi _{+}(-d_{f},\theta _{f},\omega )=0  \label{Bcf0}
\end{equation}%
guarantees the absence of a current across the insulating layer.

The boundary conditions at the FS interface strongly depend on its transport
properties. Below we will assume for the FS interface transparency $%
D(\mu_{f})\ll 1$. Then to the first approximation we may neglect the
suppression of superconductivity in the S electrode and rewrite the Zaitsev
boundary conditions \cite{Zaitsev} in the form
\begin{equation}
\frac{\ell _{f}}{k_{f}}\mu _{f}\frac{\partial }{\partial x}\Phi _{+}=D(\mu
_{f})\frac{\Delta _{0}}{\sqrt{\Delta _{0}^{2}+\omega ^{2}}}.  \label{BCSF_1}
\end{equation}

From the structure of Eilenberger equations (\ref{FplusinS})--(\ref%
{FminusinF}) and the boundary conditions (\ref{Bcds})--(\ref{BCSF_1}) it
follows that
\begin{subequations}
\begin{eqnarray}
&&\Phi _{+}(\theta _{s},\omega )=\Phi _{+}(\theta _{s},-\omega ), \\
&&\Phi_{+}(\theta _{f},\omega )=\Phi _{+}^{\ast }(\theta _{f},-\omega ),
\end{eqnarray}
\label{sym1}
\end{subequations}
\begin{subequations}
\begin{eqnarray}
&&\Phi _{-}(\theta _{s},\omega )=-\Phi _{-}(\theta _{s},-\omega ), \\
&&\Phi_{-}(\theta _{f},\omega )=-\Phi _{-}(\theta _{f},-\omega ).
\end{eqnarray}
\label{sym2}
\end{subequations}
These symmetry relations permit us to consider the solution of the boundary
problem formulated above only for $\omega >0.$

\subsection{Solution of the Eilenberger equations in the F layer.}

It is convenient to look for a solution of Eilenberger equations in the F
layer (\ref{FminusinF}) in the form
\begin{equation}
\Phi _{+}=\sum_{m=-\infty }^{\infty }Q_{m}\cos \left( \frac{\pi m(x+d_{f})}{%
d_{f}}\right) +B\cosh \left( \frac{x+d_{f}}{\ell _{f}\left\vert \mu
_{f}\right\vert }k_{f}\right) ,  \label{fi_plus_0}
\end{equation}%
which automatically satisfies the boundary condition (\ref{Bcf0}). The
relation between the coefficients $Q_{m},$ and $B$ can be found by
substitution of the ansatz (\ref{fi_plus_0}) into the Eilenberger equation (%
\ref{FplusinF}). Multiplying the obtained equations by $\cos [ \pi
k(x+d_{f})/d_{f}] ,$ and integrating them over $x$, one can easily find the
relation between the coefficients $Q_{m},$ and $B$, see the Appendix. $%
B=B(\mu_{f})$ can be found from the boundary condition (\ref{BCSF_1}). Thus
we arrive to the following expression for the Eilenberger function in the
ferromagnet
\begin{widetext}
\begin{equation}
\Phi _{+}=\frac{\Delta _{0}}{\sqrt{\Delta _{0}^{2}+\omega ^{2}}}\left[
\sum_{m=-\infty }^{\infty }\frac{\left\langle D(\mu )\frac{\mu }{q}\frac{%
\left( -1\right) ^{m}}{M(\mu )}\right\rangle _{\mu }}{M(\mu _{f})\left[
k_{f}-\left\langle \frac{1}{M}\right\rangle \right] }\cos \left( \frac{\pi
m(x+d_{f})}{d_{f}}\right) +\frac{D(\mu _{f})}{\sinh \left( \frac{q}{\mu _{f}}%
\right) }\cosh \left( q\frac{x+d_{f}}{d_{f}\mu _{f}}\right) \right] ,
\label{fi_plus}
\end{equation}%
where%
\begin{equation*}
q=\frac{d_{f}k_{f}}{\ell _{f}}=d_{f}\left( \frac{1}{\ell _{f}}+\frac{1}{\xi
_{\omega }}+i\frac{1}{\xi _{H}}\right) .
\end{equation*}%
Then the Eilenberger function averaged over the angle $\theta _{f}$ %(\ref{FplusinF})
has the form
\begin{equation}
\left\langle \Phi _{+}\right\rangle =\frac{\Delta _{0}}{\sqrt{\Delta
_{0}^{2}+\omega ^{2}}}\sum_{m=-\infty }^{\infty }\frac{\left\langle D(\mu )%
\frac{\mu }{q}\frac{\left( -1\right) ^{m}}{M(\mu )}\right\rangle _{\mu }}{%
\left[ k_{f}-\frac{q}{\pi m}\arctan \left( \frac{\pi m}{q}\right) \right] }%
\cos \left( \frac{\pi m(x+d_{f})}{d_{f}}\right) ,  \label{fi_av_dirty}
\end{equation}%

There are several interface models with a different $D(\mu _{f})$ dependence.

First, the FS interface may be represented by a thick diffusive barrier. Then the incident electron scatters in any arbitrary direction with equal probability, independent on the incident angle, i.e.,
\begin{equation}
D(\mu _{f})=D_{1}.  \label{D1}
\end{equation}

The second model considers an FS interface with a potential barrier, which appears due to different Fermi velocities in the S and F layer. The transmission coefficient in this case has the form
\begin{equation}
D(\mu _{f})=\frac{4v_{s}\mu _{s}v_{f}\mu _{f}}{\left( v_{s}\mu _{s}+v_{f}\mu
_{f}\right) ^{2}}\approx \frac{4v_{f}}{v_{s}}\mu _{f}=D_{2}\mu _{f},\quad
\text{ if}\quad v_{s}\gg v_{f}\;.  \label{D2}
\end{equation}%
It is necessary to note, that the incidence angle and the reflection angle
are related by the expression
\begin{equation}
v_{s}\sin \theta _{s}=v_{f}\sin \theta _{f}.  \label{mom_C_low}
\end{equation}%
Therefore, only the electrons which are at almost normal incidence to the interface ($\mu _{s}\approx 1 $) may penetrate through the barrier. This model seems to be mostly reasonable for
description of recent experiments \cite{Mattheiss:vF:Nb} with Nb electrodes ($v_{s}\sim 5 \ldots 6\cdot 10^{7}\,\mathrm{cm/s} $) and 3$d$ ferromagnets or
its Cu alloys \cite{Shelukhin} (for Ni $v_{f}\approx 2.8 \cdot 10^{7}\,\mathrm{cm/s} $).

Third, one can model the FS interface as a high and narrow ($\delta$-function like) potential barrier between two metals with close Fermi-velocities. In this case
\begin{equation}
D(\mu _{f})=\frac{4v_{s}\mu _{s}v_{f}\mu _{f}}{\left( v_{s}\mu _{s}+v_{f}\mu
_{f}+W\right) ^{2}}\approx \frac{4v_{s}v_{f}}{W^{2}}\mu _{f}^{2}=D_{3}\mu
_{f}^{2},\quad \text{if}\quad v_{s}\approx v_{f},  \label{D3}
\end{equation}%
where $W$ is the strength of $\delta $-function like barrier.% and $v_{s}\approx v_{f}.$

For the above mentioned three models of the FS interface we have three possible expressions for $\Phi _{+}$ with different average values $\left\langle \Phi _{+}\right\rangle $, see (\ref{D1_av})-(\ref{D3_av}) for details.

First, for $D(\mu _{f})=D_{1}$
\begin{equation}
\Phi _{+}=\frac{\Delta _{0}D_{1}}{\sqrt{\Delta _{0}^{2}+\omega ^{2}}}\left[
\sum_{m=-\infty }^{\infty }\frac{\frac{q}{2\pi ^{2}m^{2}}\left( -1\right)
^{m}\ln \left[ \frac{\pi ^{2}m^{2}}{q^{2}}+1\right] \cos \left( \frac{\pi
m(x+d_{f})}{d_{f}}\right) }{\left( m^{2}\frac{\pi ^{2}\mu _{f}^{2}}{q^{2}}%
+1\right) \left[ k_{f}-\frac{q}{\pi m}\arctan \left( \frac{\pi m}{q}\right) %
\right] }+\frac{\cosh \left( q\frac{x+d_{f}}{d_{f}\mu _{f}}\right) }{\sinh
\left( \frac{q}{\mu _{f}}\right) }\right] .  \label{Fi1}
\end{equation}%
Second, for $D(\mu _{f})=\mu _{f}D_{2}$
\begin{equation}
\Phi _{+}=\frac{\Delta _{0}D_{2}}{\sqrt{\Delta _{0}^{2}+\omega ^{2}}}\left[
\sum_{m=-\infty }^{\infty }\frac{\frac{q}{\pi ^{2}m^{2}}\left( -1\right)
^{m}\left( 1-\frac{q}{\pi m}\arctan \frac{\pi m}{q}\right) \cos \left( \frac{%
\pi m(x+d_{f})}{d_{f}}\right) }{\left( m^{2}\frac{\pi ^{2}\mu _{f}^{2}}{q^{2}%
}+1\right) \left[ k_{f}-\frac{q}{\pi m}\arctan \left( \frac{\pi m}{q}\right) %
\right] }+\frac{\mu _{f}\cosh \left( q\frac{x+d_{f}}{d_{f}\mu _{f}}\right) }{%
\sinh \left( \frac{q}{\mu _{f}}\right) }\right] .  \label{Fi2}
\end{equation}%
Third, for $D(\mu _{f})=\mu _{f}^{2}D_{3}$
\begin{equation}
\Phi _{+}=\frac{\Delta _{0}D_{3}}{\sqrt{\Delta _{0}^{2}+\omega ^{2}}}\left[
\sum_{m=-\infty }^{\infty }\frac{\frac{q}{2\pi ^{2}m^{2}}\left( -1\right)
^{m}\left( 1-\frac{q^{2}}{\pi ^{2}m^{2}}\ln \left( \frac{\pi ^{2}m^{2}}{q^{2}%
}+1\right) \right) \cos \left( \frac{\pi m(x+d_{f})}{d_{f}}\right) }{\left(
m^{2}\frac{\pi ^{2}\mu _{f}^{2}}{q^{2}}+1\right) \left[ k_{f}-\frac{q}{\pi m}%
\arctan \left( \frac{\pi m}{q}\right) \right] }+\frac{\mu _{f}^{2}\cosh
\left( q\frac{x+d_{f}}{d_{f}\mu _{f}}\right) }{\sinh \left( \frac{q}{\mu _{f}%
}\right) }\right] .  \label{Fi3}
\end{equation}%
\end{widetext}
These are the solutions of the linearized Eilenberger equation in the
ferromagnet. The corresponding expressions for $\langle \Phi _{+} \rangle$
are given by Eqs. (\ref{Fi1_av})-(\ref{Fi3_av}) in the Appendix.

Let us consider how these expressions reproduce the limiting cases of strong
and weak scattering, for which the solutions are well-known \cite%
{Buzdin:2005:Review:SF,KuprReview}.

\subsection{Dirty limit}

If the electron mean free path is the smallest characteristic length i.e. $%
\ell _{f}\ll\xi _{H},\xi _{0},d_{f},$ the frequent nonmagnetic scattering
permits averaging over the trajectories \cite{Usadel}. Then it is possible
to write a closed system of equations for averaged functions. The
linearization of the Usadel equations is allowed at the same conditions as
for the Eilenberger equation.

The linearized Usadel equation in the ferromagnet has the form
\begin{equation}
\xi _{f}^{2}\frac{\partial ^{2}}{\partial x^{2}}\left\langle \Phi
_{+}\right\rangle +\frac{\left( \omega +iH\right) }{\pi T_{c}}\left\langle
\Phi _{+}\right\rangle =0  \label{UEq}
\end{equation}%
where $\xi _{f}^{2}=D_{f}/2\pi T_{c},$ and $D_{f}=\ell _{f}v_{f}/3$ is the
diffusion coefficient. The solution of Eq.~(\ref{UEq}) may be written as
\begin{equation}
\left\langle \Phi _{+}\right\rangle =A\cosh \left( \frac{x-d_{f}}{\xi _{f}}%
\sqrt{\frac{\omega +iH}{\pi T_{c}}}\right)  \label{Uf}
\end{equation}%
The boundary condition at $x=0$ reads \cite{Kupriyanov88}%
\begin{equation}
\gamma _{B}\xi _{f}\frac{\partial }{\partial x}\left\langle \Phi
_{+}\right\rangle =\frac{\Delta _{0}}{\sqrt{\Delta _{0}^{2}+\omega ^{2}}}\;.
\end{equation}%
Substituting the expression (\ref{Uf}) into the Usadel equation (\ref{UEq})
one can find the coefficient
\begin{equation}
A=\frac{\Delta _{0}}{\gamma _{B}\sqrt{\Delta _{0}^{2}+\omega ^{2}}}\frac{1}{%
\sqrt{\frac{ \omega +iH }{\pi T_{c}}}}\frac{1}{\sinh \left( \frac{d_{f}}{\xi
_{f}}\sqrt{\frac{ \omega +iH }{\pi T_{c}}}\right)}.  \label{A}
\end{equation}
Let us consider the previously obtained averaged Eilenberger function (\ref%
{fi_av_dirty}). For small $\ell _{f}$ the parameter $q$ is large and
\begin{eqnarray*}
1 &+&\ell _{f}/\xi _{\omega }+i\ell _{f}/\xi _{H}-\frac{q}{\pi m}\arctan
\left( \frac{\pi m}{q}\right) \approx \\
&\approx &\frac{1}{3q^{2}}\left[ \pi ^{2}m^{2}+\frac{d_{f}^{2}}{\xi _{f}^{2}}%
\frac{\left( \omega +iH\right) }{\pi T_{c}}\right]
\end{eqnarray*}%
It is seen that the sums in (\ref{fi_plus}) and (\ref{fi_av_dirty}) converge
at $\pi ^{2}m^{2}\approx 2\left( \omega +iH\right) d_{f}^{2}/d_{f}. $ At
these values of $m$
\begin{equation*}
\pi ^{2}m^{2}\frac{\mu _{f}^{2}}{q^{2}}\approx \frac{2\left( \omega
+iH\right) }{d_{f}}d_{f}^{2}\frac{\mu _{f}^{2}}{d_{f}^{2}}\ell
_{f}^{2}\approx \frac{6\left( \omega +iH\right) }{v_{f}}\mu _{f}^{2}\ell
_{f}\ll 1
\end{equation*}%
and
\begin{widetext}
\begin{equation}
\Phi _{+} \approx \frac{\Delta _{0}}{\sqrt{\Delta _{0}^{2}+\omega ^{2}}}%
\left[ \sqrt{\frac{\pi T_{c}}{\omega +iH}}\frac{\cosh \left( \sqrt{\frac{\pi
T_{c}}{\omega +iH}}\frac{x+d_{f}}{\xi _{f}}\right) }{\gamma _{B}\sinh \left(
\sqrt{\frac{\pi T_{c}}{\omega +iH}}\frac{d_{f}}{\xi _{f}}\right) }+\right.
\left. \frac{D(\mu _{f})}{\sinh \left( \frac{q}{\mu _{f}}\right) }\cosh
\left( q\frac{x+d_{f}}{d_{f}\mu _{f}}\right) \right] ,
\label{fi_plus_dirty2}
\end{equation}
\end{widetext}
where the suppression parameter $\gamma _{B}$ describes the electron
transmission through the FS interface, see expressions (\ref{gammaB1})--(\ref%
{gammaB3}). After averaging over $\theta_f$ the expression (\ref%
{fi_plus_dirty2}) coincides with (\ref{Uf}) with $A$ given by (\ref{A}). It
turns out that in the dirty limit the first term of the expression (\ref%
{fi_plus}) for the Eilenberger function plays the main role, as the second
term reduces due to the spatial averaging (\ref{fi_av_dirty}), and the
Eilenberger function $\Phi _{+}=\left\langle \Phi _{+}\right\rangle $.

\subsection{Clean limit}

For a larger electron mean free path $\ell _{f}$ the sum in (\ref{fi_plus_01}%
) converges at
\begin{equation*}
\pi ^{2}m^{2}\frac{\mu _{f}^{2}}{q^{2}}\approx 1,\quad \frac{\pi m}{q}%
\approx \frac{1}{\mu _{f}}\;.
\end{equation*}%
In the limit $\xi _{H}\ll \xi _{\omega }$ we have
\begin{equation*}
k_{f}-\frac{q}{\pi m}\arctan \left( \frac{\pi m}{q}\right) \approx \ell
_{f}/\xi _{\omega }+i\ell _{f}/\xi _{H}
\end{equation*}%
and
\begin{widetext}

\begin{eqnarray}
\Phi _{+}=\frac{\Delta _{0}}{\sqrt{\Delta _{0}^{2}+\omega ^{2}}}\left\{
\left( \frac{\ell _{f}}{\xi _{\omega }}+\frac{i\ell _{f}}{\xi _{H}}\right)
^{-1}\left\langle \frac{\mu D(\mu )}{\mu _{f}^{2}-\mu ^{2}}\left[ \frac{\mu
_{f}\cosh \left( \frac{qm(x+d_{f})}{\mu _{f}d_{f}}\right) }{\sinh \frac{q}{%
\mu _{f}}}-\frac{\mu \cosh \left( \frac{qm(x+d_{f})}{\mu d_{f}}\right) }{%
\sinh \frac{q}{\mu }}\right] \right\rangle _{\mu }+\right.   \notag \\
+\left. \frac{D(\mu _{f})}{\sinh \left( \frac{q}{\mu _{f}}\right) }\cosh
\left( q\frac{x+d_{f}}{d_{f}\mu _{f}}\right) \right\} .
\end{eqnarray}%
\end{widetext}

In this limit, the first term in the square brackets in Eq.~(\ref{fi_plus})
is small in comparison with the second one. Thus, one may conclude that the
first term in Eq.~(\ref{fi_plus}) plays the main role in the dirty limit,
while the second term of Eq.~(\ref{fi_plus}) describes mainly the clean
limit when trajectories of motion are essential.

\subsection{Josephson current}

To calculate the Josephson current $J$  of a SIFS ferromagnetic tunnel
junction we start from the expression \cite{KuprReview}
\begin{eqnarray}
J &=& 4eN(0)v\pi T\sum_{\omega >0}\int_{-1}^{1}(\Imag g_{11})\cos \theta
d(\cos \theta )=  \notag \\
&=& 4eN(0)v\pi T\sum_{\omega >0}\int_{0}^{1}(\Imag g_{11}^{a})\mu d\mu .
\label{current}
\end{eqnarray}
Here $N(0)$ is the density of state at the Fermi surface, $v$ is the Fermi
velocity, $g_{11}$ is the matrix element of the antisymmetric part of the
Eilenberger function that is define by the following expression:
\begin{equation*}
\hat{g}^{a}=\frac{1}{2}\left[
\begin{array}{cc}
g(\theta )-g(-\theta ) & f(\theta )-f(-\theta ) \\
f^{+}(\theta )-f^{+}(-\theta ) & -g(\theta )+g(-\theta )%
\end{array}%
\right] .
\end{equation*}
We apply the tunnel hamiltonian approach and use the boundary conditions on
the dielectric interface. These boundary conditions are matching the
quasiclassical electron propagators $g$ and $f$ on both sides of the
boundary. The boundary conditions may essentially depend on the quality of
the interface. In the case of a nonmagnetic specularly reflecting boundary
between two metals these conditions read \cite{Zaitsev}
\begin{equation}
\hat{g}^{a}(R\hat{g}_{+}^{c2}+\hat{g}_{-}^{c2})=D_{I}\hat{g}_{-}^{c}\hat{g}%
_{+}^{c}.  \label{BG}
\end{equation}
\begin{equation}
\hat{g}_{1}^{a}=\hat{g}_{2}^{a}=\hat{g}^{a}.  \label{BGga}
\end{equation}
Here $D_{I}$ and $R_{I}=1-D_{I}$ are the interface transparency and
reflectivity coefficients, the index $1(2)$ labels the functions on the
right (left) side from the boundary plane, and the symmetric parts of the
quasiclassical Eilenberger functions are defined by the equalities
\begin{equation}
\hat{g}_{-}^{c}=\frac{1}{2}\left( \hat{g}_{1}^{c}-\hat{g}_{2}^{c}\right)
,\quad \hat{g}_{+}^{c}=\frac{1}{2}\left( \hat{g}_{1}^{c}+\hat{g}%
_{2}^{c}\right),  \label{g-g+}
\end{equation}
\begin{equation*}
\hat{g}_{1,2}^{c}=\frac{1}{2}\left(
\begin{array}{cc}
g_{1,2}(\theta )+g_{1,2}(-\theta ) & f_{1,2}(\theta )+f_{1,2}(-\theta ) \\
f_{1,2}^{+}(\theta )+f_{1,2}^{+}(-\theta ) & -g_{1,2}(\theta
)-g_{1,2}(-\theta )%
\end{array}
\right) .
\end{equation*}

Assuming that the insulating layer transparency is small, $D_{I}\ll
1,R_{I}\approx 1$, and taking into account that $\hat{g}_{1}^{2}=\hat{g}%
_{1}^{2}=\hat{1}$, we can expand the boundary conditions (\ref{BG}) in
powers of $D_{I}$.

For SIFS tunnel structures with $s$-wave pairing in the electrodes, in this
limit it immediately follows that the functions $g_{2}(\theta )=g_{s}(\theta
)$ and $f_{2}(\theta )=f_{s}(\theta )$ are independent on $\theta $ and
coincide with the expressions for spatially homogeneous superconducting
electrodes
\begin{eqnarray}
f_{f} &=& F_{f}(\theta )\exp \left\{ +i\varphi /2\right\} ,\quad
f_{s}=F_{s}\exp \left\{ -i\varphi /2\right\} ,  \notag \\
F_{s} &=& \frac{\Delta_0}{\sqrt{\omega ^{2}+\Delta_0^{2}}},\quad \quad g_{s}=%
\frac{\omega }{\sqrt{\omega ^{2}+\Delta_0 ^{2}}}\;.  \label{HomS}
\end{eqnarray}%
Here $\Delta_0$ and $\varphi $ are the absolute value and the phase
difference of the order parameter in the electrodes. The boundary conditions
(\ref{BG}) reduce in this case to
\begin{eqnarray}
\hat{g}^{a} &=& D_{I}\hat{g}_{-}^{c}\hat{g}_{+}^{c}=\frac{D_{I}}{4}(\hat{g}%
_{1}^{c}\hat{g}_{2}^{c}-\hat{g}_{2}^{c}\hat{g}_{1}^{c})=  \label{BGsDs} \\
&=& \frac{D_{I}}{4}\left(
\begin{array}{cc}
f_{1}f_{2}^{+}-f_{2}f_{1}^{+} & 2(g_{1}f_{2}-f_{1}g_{2}) \\
2(f_{1}^{+}g_{2}-g_{1}f_{2}^{+}) & f_{1}^{+}f_{2}-f_{2}^{+}f_{1}%
\end{array}%
\right) .  \notag
\end{eqnarray}
Then the expression for the Josephson current (\ref{current}) has the form
\begin{eqnarray}
J &=& \frac{1}{2}eN(0)v\pi T\sum_{\omega =-\infty }^{\infty }\int_{-\pi
/2}^{\pi /2}[f_{2}^{+}(\theta )f_{1}(\theta )-  \notag \\
&-& f_{2}(\theta )f_{1}^{+}(\theta )]\cos \theta \ d(\cos \theta ).
\label{curr1}
\end{eqnarray}
Using the definition (\ref{Fplus}) and its symmetry properties (\ref{sym1})
and (\ref{sym2}), one can find the result for the Josephson current $J$ of
the tunnel junction $J=J_{C}\sin \varphi $, where
\begin{equation}
J_{C}=\frac{8\pi T}{e R_N}\sum_{\omega >0}\Phi _{+}(\mu
_{s})\int_{0}^{1}D_{I}(\mu _{f})\Real\left[ \Phi _{+}(\mu _{f})\right] \mu
_{f}d\mu _{f} \;.  \label{Jc}
\end{equation}
The Eilenberger function of the left electrode $\Phi _{+}(\mu _{s})$ is also
defined by (\ref{Bcds}).

The thin insulating layer is considered as a high potential barrier for
electrons. The transmission probability is inversely proportional to the
exponent of a distance passed by an electron, i.e.~$\sim \exp(-d_{I}/\cos
\theta ),$ where $d_{I}$ is the dielectric thickness. Namely, it can be
found from the well known expression \cite{LLQmech} for the transmission
coefficient of a square potential barrier of a thickness $a$, that
\begin{eqnarray}
D_{I}(a) &=&\frac{4k_{1}^{2}k_{2}^{2}}{%
4k_{1}^{2}k_{2}^{2}+(k_{1}^{2}+k_{2}^{2})^{2}\sinh ^{2}(k_{2}a)}\approx
\notag \\
&\approx &\frac{4k_{1}^{2}k_{2}^{2}}{(k_{1}^{2}+k_{2}^{2})^{2}}\exp
(-2k_{2}a),
\end{eqnarray}%
if $k_{2}a\gg 1$. Here $k_{1}$ and $ik_{2}$ are wave vectors of particles
outside and inside the barrier. The expressions for the transmission
coefficient of the FS boundary (\ref{D2}),(\ref{D3}) were obtained in the
same way. Taking into account that $a=d_{I}/\mu ,$ one can write the
dependence $D_{I}(\mu )$ in the form
\begin{equation}
D_{I}(\mu )=\widetilde{D_{0}}\exp \left( -\frac{\alpha }{\mu }\right) .
\end{equation}%
Here $\alpha $ is a decaying coefficient, that depends on the thickness and
material of the insulator. It is useful for a numerical calculation to
redefine the value $\widetilde{D_{0}}=D_{0}\exp (-\alpha )$ , and finally
\begin{equation}
D_{I}(\mu )=D_{0}\exp \left( -\alpha \frac{1-\mu }{\mu }\right) .  \label{D0}
\end{equation}

\section{Discussion}

\label{Sec:Discussion}

\subsection{Main cases}

First, we start from the analysis of the Eilenberger function $\Phi_+ (x,
\mu_f)$, which describes the superconducting properties of a ferromagnetic
Josephson junction. In the simplest case $\Phi_+ (x, \mu_f)$ is an
exponential function $\Phi_+ (x)\propto \exp (x/\xi ), \ x\leqslant 0$, or a
combination of exponential functions, with a complex coherence length \cite%
{Buzdin:2005:Review:SF,KuprReview}
\begin{equation}
\xi^{-1} = \xi_1^{-1}+i \xi_2^{-1} .  \label{ksi}
\end{equation}
Here $\xi_1$ describes the decay of the superconducting correlations at some
distance from the FS boundary, while $\xi_2$ defines the period of LOFF
oscillations \cite{LO,FF}. For the described FS structure, $\Phi_+ (x, \mu_f)
$ is given by the expressions (\ref{fi_plus}) and (\ref{Fi1})-(\ref{Fi3}).
Now we try to describe the main features of the proximity effect in an FS
structure. If the superconducting layer is thick enough to be considered as
semi-infinite, and if in-plane nonuniformities connected with sample
preparation and the domain structure of the ferromagnetic film can be
neglected, the structure has one geometric parameter --- the ferromagnet
thickness $d_f$. The properties of the F material give three more
characteristic lengths: the electron mean free path in the ferromagnet $\ell
_{f}$, the nonmagnetic coherence length $\xi_{0}$, and the characteristic
magnetic length $\xi _{H}$. The main questions at this stage are: can the
Eilenberger function be approximated by an exponential function, and what is
the relation between $\xi_1, \xi_2$ and the characteristic lengths mentioned
above.

The Fermi velocity for usual ferromagnetic metals \cite{Shelukhin} is $%
v_{f}\sim 3 \cdot 10^7 \,\mathrm{cm/s} $, and if even the temperature $T
\sim T_c$, $\xi_0 \sim 100 \,\mathrm{nm} $, i.e.~$\xi_0$ is much larger than
other parameters of the problem in usual cases. The value of $\xi_0$ is
difficult to decrease. The gradual increase of $\ell_f$ leads to the
following four cases:

1. $\ell_f \ll \xi_H, d_f$ --- the well investigated dirty limit. This
condition allows averaging of the Eilenberger functions over trajectories
and using the Usadel equations. $\Phi_+$ is given by the expression (\ref{Uf}%
), and for a strong exchange field ($H\gg k_B T_c$), $\xi_1= \xi_2 = \sqrt{2
\ell_f \xi_H /3} $.

2. $\ell_f \sim \xi_H, d_f$ --- the intermediate case. The Usadel equations
cannot be used, and one needs to solve the Eilenberger equations taking into
account the $\Phi_+ (\theta)$ dependence. Up to now the linearized equations
were solved only for an SFS junction \cite{Volkov,Linder}, and the obtained
expressions are so complicated that they are very difficult to analyze. This
is the most interesting case for our analysis, which allows to find out when
this case reduces to the dirty limit and under which conditions the Usadel
equations are applicable.

3. $\xi_H \ll \ell_f <d_f, \xi_0$ --- the clean limit. Here, $\Phi_+ (x,
\theta)$ is defined mainly by the second term of Eq.~(\ref{fi_plus}), and
\cite{Volkov,OurEuLett} $\xi_1 \sim \ell_f$, $\xi_2 \approx \xi_H $.

4. $\ell_f \gg d_f$ --- the ballistic regime cannot be considered within the
framework of our approach, excluding the case $\xi_1 \ll d_f$ when multiple
reflections from the interfaces can also be neglected. The ballistic SFS
junction was considered earlier\cite%
{Buzdin:1982,Radovic:2001:0&pi-JJ,Buzdin2-3D,Linder,Radovic2003,Zareyan1},
and the dependence $J_C (d_f)$ cannot be presented as an exponent in the
general case \cite{Radovic2003}.

\subsection{Analysis of cases 1-3}

In all plots presented below we use the same normalization for all lengths: $%
x$, $d_f$, $l_f$, $\xi_0$, $\xi_H$, $\xi_1$, $\xi_2$. All of them are
normalized to some unit length $\Xi$, which, in fact, can be chosen
arbitrarily, e.g. $\Xi=1\,\mathrm{nm} $. Note that only the ratios between
the lengths given above are important, i.e., if one divides the above list
of lengths by any arbitrary constant $\Xi$ all the results remain unchanged.
For example, in all plots we use $\xi_0=81$.

\paragraph{Influence of different types of FS boundary transparency.}

\begin{figure}[!htb]
\begin{center}
\includegraphics{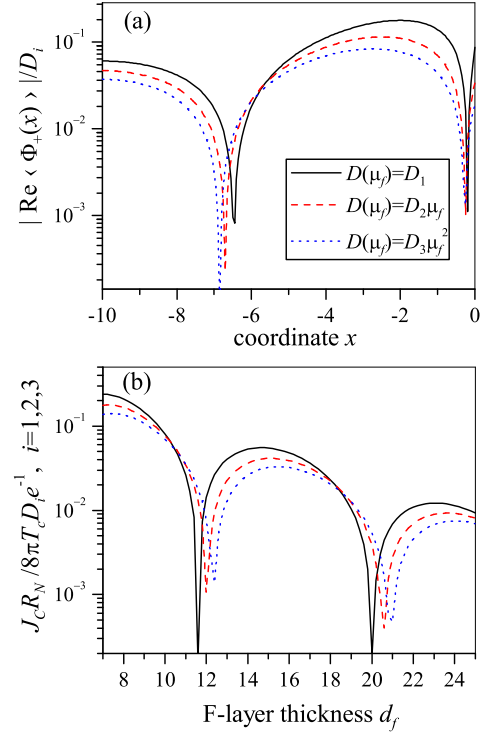}
\end{center}
\caption{(Color online) (a) Averaged Eilenberger function $\langle\Phi_+
(x)\rangle$ and (b) $J_C (d_f) R_N$ product for different types of the FS
interface (\protect\ref{D1})-(\protect\ref{D3}). The F-layer parameters are $%
\ell_f=7,\ \protect\xi_H=3$ (a,b), $d_f=10$ (a), the insulating layer decay
parameter is $\protect\alpha=5$ (b); $T\approx0.5T_c$. }
\label{fig:D1-D3}
\end{figure}

Figure~\ref{fig:D1-D3} shows the spatial distribution of the functions $\langle\Phi_+ (x)\rangle$ calculated from (\ref{Fi_av}) for
$\ell_f=7,\ \xi_H=3$, $d_f=10$
and $J_C (d_f)$ calculated from (\ref{Jc}) for $\ell_f=7,\ \xi_H=3$, and the insulating layer decay parameter $\alpha=5$.
It is clearly seen that the angular dependence of SF interface transparency influences both, the amplitude and period of oscillations of the presented curves.
For the particular choice of parameters this influence can be easily explained.

For the case of a transparency which is independent on $\mu_f$, the Eilenberger functions are initiated from the SF interface in all directions, thus leading to the largest amplitude of  $\langle\Phi_+ (x)\rangle$ oscillations. Simultaneously, the contributions to $\langle\Phi_+ (x)\rangle$ coming from rapidly decaying  Eilenberger functions with a large argument $\theta _{f}$ result in the appearance of the smallest period of $\langle\Phi_+ (x)\rangle$ oscillations. The stronger the angular dependence of the SF interface transparency coefficient, the smaller are the contributions to average values from the rapidly decaying Eilenberger functions, and as the result the larger is the period and the smaller is the amplitude of   $\langle\Phi_+ (x)\rangle$ oscillations. It is necessary also to mention that $\langle\Phi_+ (x)\rangle$ decays more rapidly with $|x|$ than $\Phi_+ (x,1)$. This difference is the smaller the smaller is $l_{f}.$

From (\ref{Jc}) it follows that in general the contribution to the junction critical current comes not only from $\Phi_+ (x,1)$, but also from $\Phi_+ (x,\mu_f)$ located in a narrow domain of $\mu_f$ nearby $\mu_f = 1.$ Due to that, the  $J_{C}(d_{f})$ curve appears to be sensitive against the angular dependence of the SF interface transparency. We see that the stronger this dependence the smaller is the amplitude of oscillations and the larger are the distances of the $0$ to $\pi$ transition points in $J_{C}(d_{f})$ from the SF interface, while the period of $J_{C}(d_{f})$ oscillations is practically  insensitive against the form of $D(\mu_{f})$ and coincides with that of $\Phi_+ (x,1).$
If the ferromagnetic layer is thick enough, the sum over $\omega $ in (\ref{Jc}) converges rapidly, and the main contribution is given by the first term of the sum, which is determined by the real part of the Eilenberger function $\Phi_+ (x,1)$.
For these reasons we will focus below on the examination of the properties of the real part of the Eilenberger function $\Phi_+ (x,1)$, and in further calculations we will also use the form $D(\mu_f)=D_2 \mu_{f}$ for the transmission coefficient of the FS interface, as it is most applicable to materials used in experiments.

\paragraph{Spatial profile of the Eilenberger function.}

\begin{figure}[!htb]
\begin{center}
\includegraphics{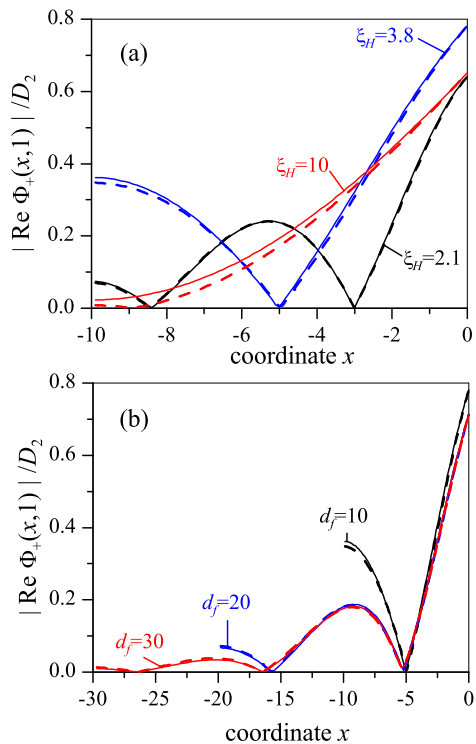}
\end{center}
\caption{(Color online) The exact $\Phi_+ (x,1)$ (solid lines) and
approximated $\Phi_+ (x,1)^{app}$ (dashed lines) Eilenberger function (a)
for different values of $\protect\xi_H$ corresponding to a local maximum,
minimum, and smooth region of the dependence $\protect\xi_1 (\protect\xi_H)$%
; see Fig.\protect\ref{fig:xi12(xiH)}(b), at $\ell_f=7,\ d_f=10$; (b) for
different thicknesses $d_f$ at $\ell_f=7,\ \protect\xi_H=3.8$. $%
T\approx0.5T_c$. }
\label{fig:FiAppr}
\end{figure}

The direct comparison (see Fig. \ref{fig:FiAppr}) of the results of numerical calculations of the real part of the Eilenberger function $\Phi_+ (x,1)$, making use of Eq.~(\ref{Fi2}), with the curves which follow from the simple analytical expression
\begin{equation}
\Phi _{+}^{app}(x,1)\approx \frac{\Phi _{+}(0,1)}{\cosh (d_{f}/\xi )}\cosh
\frac{x+d_{f}}{\xi }  \label{Fi_appr}
\end{equation}
confirms that it is possible to approximate $\Phi_+ (x,1)$ by this simple formula. It satisfies the boundary condition on the dielectric interface (\ref{Bcf0}) and contains the factor $\Phi _{+}(0,1)$, which has been calculated numerically from (\ref{Fi2}).

The quality of the approximation (\ref{Fi_appr}) was
checked for different values of parameters for all dependencies investigated
further. It was found to be satisfactory in the whole accessible range of
parameters $d_f>\ell_f \ \mathrm{\ or }\ \xi_1$. This approximation yields
the values of $\xi$, i.e. $\xi_1\ \mathrm{\ and }\ \xi_2$.

\paragraph{Dependence of $\protect\xi$ on the junction parameters.}

The attempt to find $\xi$ in the area of intermediate scattering was made
earlier \cite{Gusakova}. It was supposed in Ref.~[\onlinecite{Gusakova}] that $\xi$ may be found as a
solution of the equation
\begin{equation}
 i \ell_f/ \xi=\arctan(i \ell_f/ k_f\xi), \label{gus}
\end{equation}

 which determines the poles of the sum in Eqs.(\ref{fi_plus}), (\ref{fi_av_dirty})
 for $\Phi_+ (x)$ and $\langle\Phi_+ (x)\rangle,$ respectively. Unfortunately, the inverse tangent is a multivalent function. In the general case this fact prevents
 to represent $\Phi_+ (x)$ and $\langle\Phi_+ (x)\rangle$ as only the sum of residues of summable functions in (\ref{fi_plus}), (\ref{fi_av_dirty}). An exception occurs in the range of parameters for which the ferromagnet is close to the dirty limit.  In this case the sum in (\ref{fi_plus}) and (\ref{fi_av_dirty}) converts faster than the multivalent nature of the inverse tangent becomes essential. These difficulties had been first pointed out in Ref.~[\onlinecite{VolkovCritKupr}], where it was  also demonstrated that the solution of the Eilenberger equations used in Ref.~[\onlinecite{Gusakova}] does not transfer correctly to the solution of the Eilenberger equations in the clean limit.

\begin{figure}[!htb]
\begin{center}
\includegraphics{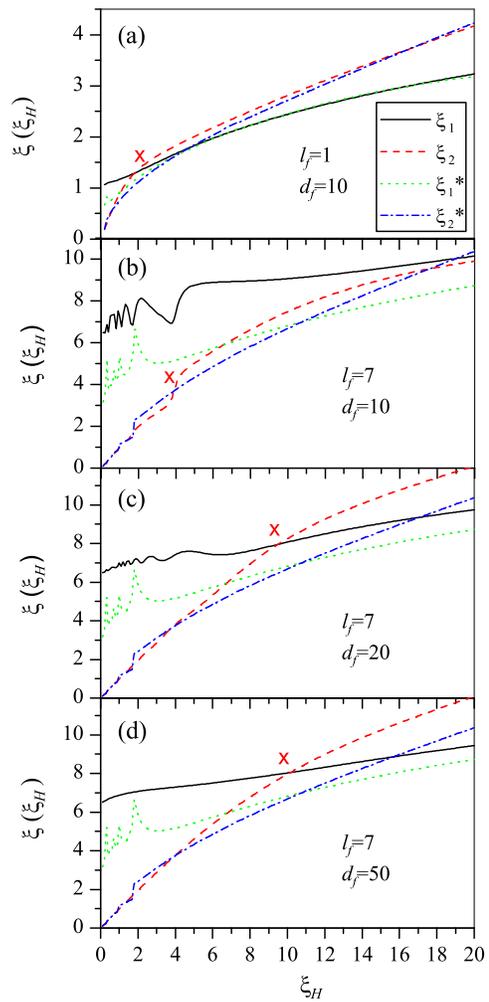}
\end{center}
\caption{(Color online) Decay length $\protect\xi_1$ and oscillation
parameter $\protect\xi_2$ vs magnetic length $\protect\xi_H$ for different
values of $\ell_f$ and $d_f$, that are denoted in the plots. The
corresponding values of $\protect\xi$ from Ref.~[\onlinecite{Gusakova}] are
also shown and marked by $*$. }
\label{fig:xi12(xiH)}
\end{figure}

To avoid these mathematical difficulties we have decided to find out the dependence $\xi (\xi_H,\ell_f)$
from the direct fitting of the real part of $\Phi_+ (x)$ calculated numerically from (\ref{fi_plus}) by approximating formula (\ref{Fi_appr}).
The result of this procedure is demonstrated in Fig. \ref{fig:xi12(xiH)}. It
shows the dependencies of $\xi_1 \mathrm{\ and }\
\xi_2$ on the magnetic length $\xi_H$ at different $\ell_f$. The solutions
from Ref.~[\onlinecite{Gusakova}] are also presented in Fig.~\ref%
{fig:xi12(xiH)} for comparison. They coincide with our values when $%
\xi_H \gg \ell_f$ i.e. in the dirty limit. We have taken two cases for the
numerical investigation: $\ell_f=1$, that is close to the dirty limit, and $%
\ell_f=7$, that is close to the clean case.

\begin{figure}[!htb]
\begin{center}
\includegraphics{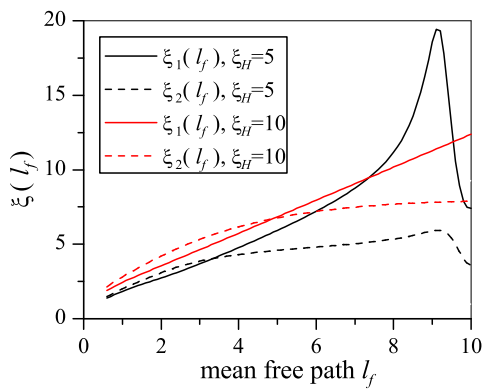}
\end{center}
\caption{(Color online) Decay length $\protect\xi_1$ (solid lines) and
oscillation parameter $\protect\xi_2$ (dashed lines) vs electron mean free
path $\ell_f$ in the ferromagnet for different values of $\protect\xi_H$;
here $d_f=10$, $T\approx0.5T_c$. }
\label{fig:ksi(lf)}
\end{figure}

It is seen, that $\xi_1$ oscillates with the exchange energy $H$ (so as $%
\xi_H \sim H^{-1}$), an interesting result, that was not noted earlier.
These oscillations appear in rather clean samples $\ell_f\gtrsim d_f/ \pi$
and for $\xi_H < d_f/ \pi$. They originate mainly from the second term of
the expression (\ref{fi_plus}), which plays the main role in the clean case.
The function $\xi (\xi_H)$ changes with increasing $d_f$ (cf.~Fig.~\ref%
{fig:xi12(xiH)}(b--d)). At large enough $d_f$, electrons have a time to
scatter in the ferromagnet, their trajectories shuffle, and the dominant
role of the ''clean`` term vanishes, the oscillations disappear, see Fig.~\ref%
{fig:xi12(xiH)}(d).
We note that the smaller is the probability of dephasing of
the Eilenberger functions due to electron scattering in the F layer,
the stronger are the interference effects between the IF and FS interfaces.
This interference demands, for given values of $\xi_1$ and $\xi_2$,
the fulfillment of the boundary conditions at these interfaces,
which fix the value of the phase derivatives at $x=0$ and $x=d_f$.
As it follows from Eq.~(\ref{Fi1})-(\ref{Fi3}), these conditions cannot be satisfied
on the class of monotonic $\xi_1$ and $\xi_2$ functions.
Therefore the transition from the dirty to the clean limit should be
accompanied by a nonmonotonic behavior of $\xi_1$ and $\xi_2$ as
a function of $\xi_H$ for fixed $\ell$, see Fig.~\ref{fig:xi12(xiH)},
or as a function of $\ell$ for fixed $\xi_H$, see Fig.~\ref{fig:ksi(lf)}.

In the dirty limit the spatial oscillation length $\xi_2\approx \sqrt{%
2\ell_f \xi_H /3}$, while in the clean limit $\xi_2\approx \xi_H$. Near the
point marked by X in Fig.~\ref{fig:xi12(xiH)} there is a crossover between the dependencies corresponding to dirty and clean limits, respectively, i.e.~this region can be considered as a boundary between the clean
and the dirty cases.

%\label{fig:ksi(lf)}

The dependence of $\xi$ on the mean free path $\ell$ for different $\xi_H$
is presented in Fig.~\ref{fig:ksi(lf)}. Usually the decay length $\xi_1 $ as
well as the oscillation period $2\pi \xi_2$ increase with the mean free
path, which corresponds to the results in Ref.~[\onlinecite{Linder}].
However, the dependence $\xi(\ell_f)$ may be nonmonotonic for some values of
the ferromagnet thickness.

In the presented figures one can see, that $\xi_1$ may be larger or smaller
than $\xi_2$ and may behave nonmonotonically. This depends not only on the material constants of
ferromagnetic material, but also on the thickness of the F layer.

The found values $%
\xi_1\ \mathrm{\ and }\ \xi_2$ together with the simple dependence (\ref%
{Fi_appr}) could be widely used for various estimations, approximate
calculations, and fitting experimental data, for example for a measurement of
the density of states (DOS) in an FS bilayer or calculation of thickness dependence of the critical current of SIFS Josephson junctions.

\paragraph{$J_C (d_f)$ in the limit of large $d_f$.}

As an example, we consider the critical current of SIFS Josephson junctions in the limit of large F layer thickness $d_f \gg \xi_{1}$.  In this limit the main contribution to the critical current in (\ref{Jc}) is given by the term at  $\omega=\pi T.$ Suppose further that only  electrons, that are  incident in the direction perpendicular to the FI interface, provide the current across it, from (\ref{Jc}), (\ref{Fi_appr}) we have
\begin{equation}
J_{C}=\frac{8\pi TD_{2}}{eR_{N}}\Real{\frac{\Delta _{0}\Phi _{+}(0,1)}{\sqrt{(\pi
T)^{2}+\Delta _{0}^{2}}\cosh (d_{f}/\xi )}} \label{icass}
\end{equation}

\begin{figure}[!htb]
\begin{center}
\includegraphics{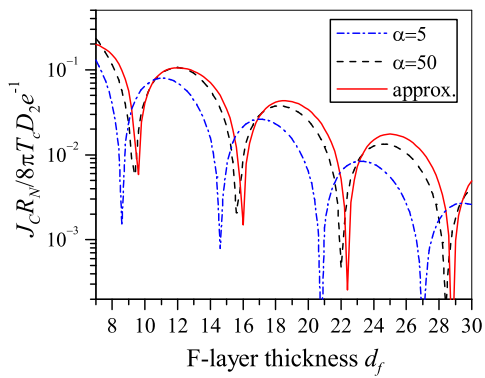}
\end{center}
\caption{(Color online) $J_C R_N$ product as a function of the ferromagnet
thickness $d_f$ at different insulating layer thicknesses described by the
parameter $\protect\alpha$, and the approximation by expressions (\protect
\ref{Jc}),(\protect\ref{Fi_appr}) at the value $\protect\xi$ taken at $d_f=50
$, see Fig.\protect\ref{fig:xi12(xiH)}(d), here $l_f=7$, $\protect\xi_H=2.1$%
, $T\approx0.5 T_c$,. }
\label{fig:Jc_alpha}
\end{figure}

The comparison with the exact numerical calculation has shown that this expression (\ref{icass}) well approximates $J_C (d_f)$ starting from $d_f \gtrsim \xi_{1}$.
The thickness dependence of $J_C$ calculated from (\ref{icass}) at the value $\xi$ taken at $d_f=50
$, see Fig.\protect\ref{fig:xi12(xiH)}(d), for $l_f=7$, $\xi_H=2.1$%
, $T\approx0.5 T_c$ is shown in Fig. \ref{fig:Jc_alpha} by the solid line.
The dashed and dashed dotted curves in Fig. \ref{fig:Jc_alpha} give the results obtained by numerical calculation, which have been done for the same parameters with the use of  the exact expression for $%
\Phi_+ (-d_f,\mu_f)$  in (\ref{Jc}) and at different insulating layer thicknesses described by the
parameter $\alpha.$

It is clearly seen that the larger is $\alpha$ the closer is the the result to the approximation formula (\ref{icass}). We may also conclude that the difference between the asymptotic solid curve and the dashed  curves in Fig. \ref{fig:Jc_alpha} calculated for finite values of $\alpha$ occurs only in amplitude and positions of the $0$ to $\pi$ transition points, while the decay length and period of  $J_C(d_{f})$ oscillations are nearly the same. This means, that if in experiment we are mainly interested in the estimation of the ferromagnet material constant, $\xi,$ we may really use for the data interpretation the simple expression (\ref{icass}) and consider the coefficient $\Phi _{+}(0,1)$ in it as a phenomenological parameter, which depends not only on the properties of ferromagnetic material, but
also on $d_f$ in a rather complicated way, see Fig.\ref{fig:FiAppr}(b), as well as on the form of the transparency coefficients at the FI and FS interfaces.
Although our results for SIFS junctions give the same
$\xi_1$ and $\xi_2$ as for SFS junctions (at least for a thick F layer),
different boundary conditions result in different order parameter amplitude and phase.
This results in a different $J_C (d_f)$ dependence, similar to the
results obtained earlier in dirty limit \cite{Vasenko}.

It is interesting to note that in the case of a rather thin insulating
barrier and rather clean ferromagnet, $\xi_1\ \mathrm{\ and }\ \xi_2$
as measured from the critical current of SIFS junctions may differ from $\xi$
as measured from the DOS on the free F surface of an analogous FS bilayer. This is
because the DOS measured by low-temperature scanning tunneling spectroscopy (which may only yield contributions from electrons with normal incidence) is directly given by $\Phi_+(-d_f,\mu_f=1)$, while $J_C$ includes contributions from other angles.

\begin{figure}[!htb]
\begin{center}
\includegraphics{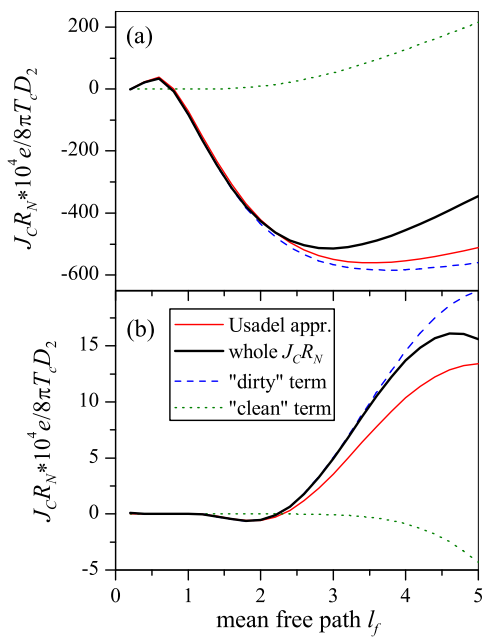}
\end{center}
\caption{(Color online) $J_C R_N$ product vs ferromagnet mean free path $l_f$
($\protect\xi_H=10$, $T\approx0.5 T_c$, $\protect\alpha=5$) for $d_f=10$ (a)
and $d_f=30$ (b). Each plot shows $J_C R_N$ obtained from the Usadel
function (\protect\ref{fi_av_dirty}), from the Eilenberger function $\Phi_+$
(\protect\ref{Fi2}), as well as from only the first "dirty" term or only the
second "clean" term of (\protect\ref{Fi2}). }
\label{fig:limUsadel}
\end{figure}

\paragraph{Applicability of the Usadel equation.}

The condition that allows using the Usadel equation is well known --- this
is strong nonmagnetic scattering \cite{Usadel}, namely, $\ell_f \ll \xi_H,
d_f, \xi_0$. What does the symbol ''$\ll$`` mean exactly? In order to
illustrate this, Fig.~\ref{fig:limUsadel} shows the dependence $J_C(\ell_f)$
given by the expression (\ref{Fi2}). Two parts of (\ref{Fi2}), called for
convenience "dirty" and "clean", are given by the first and the second terms
of the expression (\ref{Fi2}). The critical current density as given by the
Usadel function (\ref{fi_plus_dirty2}), describing the dirty limit, is also
shown. $J_C(\ell_f)$ coincides with the Usadel solution at $\ell_f\approx
0.1 \xi_H$, then the first term in square brackets in Eq. (\ref{Fi2})
dominates and the $\mu_f$-dependence becomes nonsignificant due to the spatial
averaging as a result of multiple scattering. This means, that the Usadel
equations become appropriate to use if the parameter $H\tau_f \leqslant 0.1$%
. %where $\tau=\ell_f / v_f$ is mean free path.
This result was intuitively clear, but demanded a proof due to plenty of
investigations, where the Usadel equations was used at
$H\tau_f \lesssim 1$.

\section{Comparison with experiment}

\label{Sec:Exper}

There are only few experiments on SIFS tunnel junctions with moderate
scattering in the F-layer \cite{Born:2006:SIFS-Ni3Al,WeidesCleanSIFS}. In
Ref. [\onlinecite{Born:2006:SIFS-Ni3Al}] experimental data points for $J_C
(d_f)$ are rather sparse, while in Ref. [\onlinecite{WeidesCleanSIFS}] the
density of data points per period is much larger. We thus decided to compare
our theory to the data of Ref. [\onlinecite{WeidesCleanSIFS}].

This article \cite{WeidesCleanSIFS} also contains attempts to fit the
experimental data by theoretical curves, however, without much success. Some
of these attempts, see Fig.5(a) in Ref.~[\onlinecite{WeidesCleanSIFS}], use
a theory developed for a very clean (ballistic) SFS junction. The predicted
oscillation period of $J_C(d_f)$ was significantly smaller than the one in
the experiment. Figure 5(b) in Ref.~[\onlinecite{WeidesCleanSIFS}] shows the
same experimental data together with fits using dirty limit theory. The best
fit was achieved with $\xi_1=0.66\,\mathrm{nm} $ and $\xi_2=0.53\,\mathrm{nm}
$. However, the theory for a dirty ferromagnetic junction cannot explain $%
\xi_1>\xi_2$, and all of these theories, both clean and dirty, do not take
into account the insulating layer.

In Ref. [\onlinecite{WeidesCleanSIFS}] and Ref. [\onlinecite{WeidesAnisotr}]
it was found, that the magnetic anisotropy of the F layer changes from
perpendicular to in-plane with an increase of the F-layer thickness.
Perpendicular magnetic anisotropy in a polycrystalline thin film may occur
due to several reasons: the mechanism described by Neel \cite{Neel} related
to the absence of nearest neighbor F-layer atoms near interfaces, the
magneto-striction mechanism, the surface roughness, and the one associated
with microscopic-shape anisotropy \cite{FilmHandbook}. The latter may also
vary for samples sputtered under different angles \cite{FilmHandbook}.
Perpendicular magnetic anisotropy can change to an in-plane anisotropy due
to a competition with the shape anisotropy of a film. However, none of the
mechanisms listed above can be expected to give a significant jump of the
exchange magnetic energy $H$ at the transition between the two types of
anisotropy. There is some shift of the experimental curve, see Fig.2(a) in
Ref.~[\onlinecite{WeidesCleanSIFS}], at $d_f=3.8 \,\mathrm{nm} $, where the
anisotropy changes. But there are no reasons to assume, that this change of
anisotropy is associated with a significant change of $H$. Therefore, we try
to describe the whole experimental curve $J_C(d_f)$ using the same value of $H$.
%, which for $1\,\mathrm{nm} \leqslant d_f \leqslant 3.8\,\mathrm{nm} $ is
%fitted with larger accuracy than for $d_f>3.8\,\mathrm{nm} $ .

The common problem for all explanations of this experiment is the location
of the first minimum of $J_C(d_f)$ at a rather large $d_f\approx3\,\mathrm{nm%
} $ in comparison with the oscillation length $\xi_2 \approx 0.5\,\mathrm{nm}
$. The authors \cite{WeidesCleanSIFS} attribute this to the existence of a
nonmagnetic dead layer with rather large thickness $d_{dead}=2.26\,\mathrm{nm%
} $ in the Ni layer.

We note, that $d_{dead}$, in general, depends on $d_f$. We assume that $%
d_{dead}=d_f$ up to some value $d_{max}$ and then either $d_{dead}$ stays
constant at a saturation value $d_{dead}^{sat}=d_{max}$ with further
increase of $d_f$, or it is reduced to a constant value $%
d_{dead}^{sat}<d_{max}$. For the latter case there are the following
reasons. The dead layer may consist of a nonmagnetic Ni alloy with magnetic
clusters or islands, which are not connected by the exchange interaction. A
very thin film of a ferromagnetic material may be paramagnetic. When one
covers it by extra magnetic layers, some part of its thickness may
magnetize. First, magnetic clusters uncoupled by the exchange interaction
can couple if they are covered by extra magnetic layers. Second, a thin
homogeneous film of a magnetic material of the thickness of a few monolayers
exhibits nonmagnetic properties alone, but becomes magnetic entirely when
its thickness increases, due to the transition from the two-dimensional to
the three-dimensional case.

The dead layer is described as a normal layer, that yields $J_{C}(d_{f})\sim
\exp (-\frac{d_{f}}{\xi _{N}})$. The best fitting presented in Fig.~\ref%
{fig:exper}(a),(b) by straight short-dashed lines, gives the value\cite%
{WeidesCleanSIFS} $\xi _{N}=0.68\,\mathrm{nm}$. The normal metal coherence
length $\xi _{N}$ depends both on the nonmagnetic and paramagnetic scattering.
The latter may be significant in the dead layer in the presence of many magnetic inhomogeneities (clusters and impurities). Following a theory
developed for SFS junctions\cite{RyazanovBuzdinJc(d)PRL2006} taking into
account magnetic scattering, one can find the expression for the
coherence length of a normal metal (at $H=0$) with magnetic impurities $\xi _{N}=\sqrt{\frac{D_{f}}{2\left( \pi T+1/\tau _{m}\right) }}$,
where $\tau _{m}$ is the magnetic scattering time. We assume that the mean
free path has the same value as in the rest of the magnetic Ni layer, $\ell
_{f}=0.6\,\mathrm{nm}$, that is given by the fit. This yields
$\tau _{m}=1.68\cdot 10^{-14}\,\mathrm{s}$. For comparison, the
nonmagnetic scattering time is found to be almost 8 times less. Fig.~\ref%
{fig:exper}(a),(b) shows that the obtained value of $\xi _{N}$ is very close
to the decay length of the magnetic SIFS junction at $d_{f}>d_{max}$. In the
absence of the exchange field its pair-breaking role is played by the
magnetic scattering in this structure.

\begin{figure}[!htb]
\begin{center}
\includegraphics{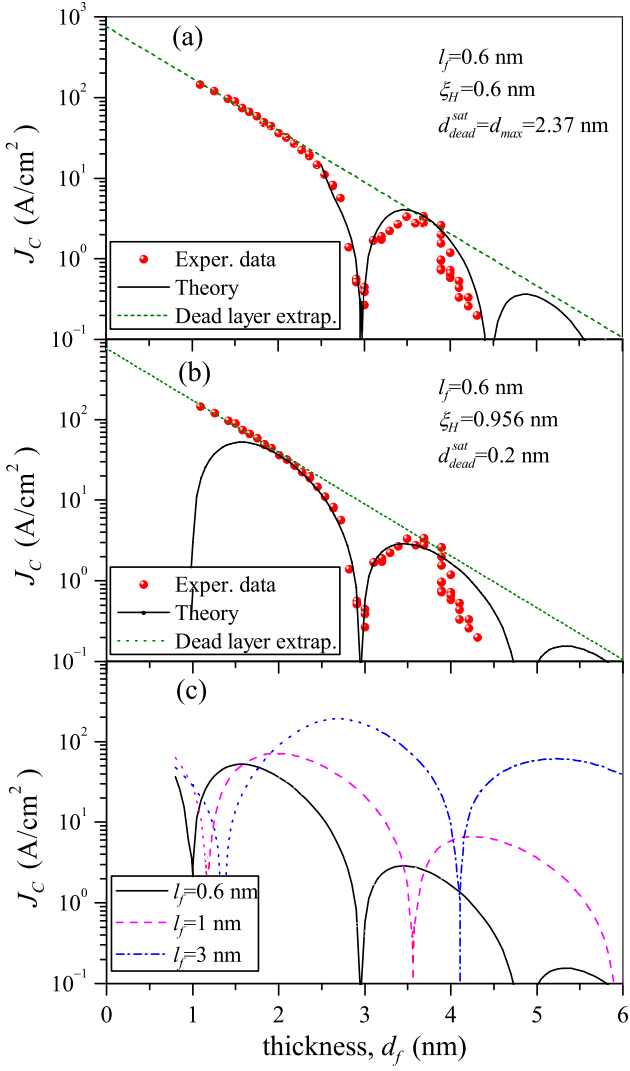}
\end{center}
\caption{(Color online) (a) and (b) show a comparison of $J_C (d_f)$ from the presented theory with experiment \protect\cite{WeidesCleanSIFS}.
Dashed lines are fits to the experimental data in the range $d_f<2.37\units{nm}$, where the F layer can be interpreted as nonmagnetic.
(c) Dependencies $J_C (d_f)$ at lower scattering. Parameters
are like in (b). The regime for which multiple reflections cannot be
neglected are marked by dotted lines.
}
\label{fig:exper}
\end{figure}

We consider two possibilities to explain the experimental data \cite%
{WeidesCleanSIFS}. First, we assume that the detected minimum of $J_C (d_f)$
is the first one, see Fig.~\ref{fig:exper}(a). Treating the dead layer as a
normal layer ($H=0$), our best fit yields $d_{dead}^{sat}=d_{max}=2.37\,%
\mathrm{nm} $. The fitting yields the following values of the F-layer
parameters: $\ell_f=0.6\,\mathrm{nm} $ and $\xi_H=0.6 \,\mathrm{nm}$, that
corresponds to $H=1782\,\mathrm{K}$ or $H=154\,\mathrm{meV}$.

As a second scenario we assume that the detected minimum of $J_C(d_f)$ is
the second one, see Fig.~\ref{fig:exper}(b). The first $J_C(d_f)$ minimum
may not show up if the effective thickness of the dead layer \textit{%
decreases} as soon as $d_f$ is above the threshold value $d_{max}$. Then, at
the value of $d_f$ where
$J_C(d_f)$ is supposed to have the first minimum, the F layer may still be nonmagnetic,
while at large values of $d_f$ it has some small value $d_{dead}^{sat}$. For
that scenario $d_{dead}^{sat}=0.2\,\mathrm{nm} $ provides the best fit. The
mean free path was found as $\ell_f=0.6\,\mathrm{nm} $, $\xi_H=0.956\,%
\mathrm{nm} $, that corresponds to $H=1119\,\mathrm{K}$ or $H=96.4\,\mathrm{%
meV}$.

In both fits we used the following parameters. The temperature $T=4.2\,%
\mathrm{K}$ and $T_c=8\,\mathrm{K} $ were taken like in the experiment\cite%
{WeidesCleanSIFS}, that yields $\xi_0=81 \,\mathrm{nm} $. Fermi velocities
\cite{Shelukhin} $v_{f}=2.8\cdot 10^{7}\,\mathrm{cm/s}$ and \cite%
{Mattheiss:vF:Nb} $v_{s}=6.2\cdot 10^{7}\,\mathrm{cm/s}$ were taken. Since $%
v_{s}>v_f$, we use the FS boundary condition in the form (\ref{D2}). The
thickness of the insulating $\mathrm{Al_2O_3}$ barrier was not measured
exactly; still the value $\alpha =5$ seems to be realistic, see Eq.(\ref{D0}%
). The values of the exchange field for both cases are within the range
obtained for Ni in SFS experiments\cite%
{RobinsonPRB76,RobinsonPRL97,Blum:2002:IcOscillations,Shelukhin}. The
relation $\ell_f \sim \xi_H$ does not allow using the Usadel equations,
however, this is not far from the dirty limit.

Comparing both fits shown in Fig.~\ref{fig:exper}(a),(b) we cannot give a
definite answer which minimum (the first or the second) was observed in the
experiment \cite{WeidesCleanSIFS}, but we are able to explain the fact that $%
\xi_1>\xi_2$ in any case. Such relation between $\xi_1$ and $\xi_2$ cannot
be obtained from a dirty limit theory.

%\label{fig:exper}

The Josephson junctions used in the experiment \cite{WeidesCleanSIFS}
contain an extra normal layer of Cu between the I and F layer. A normal
layer cannot change $\xi$, which mainly depends on properties of the
magnetic layer. However, it may change the boundary conditions. Therefore,
our theory cannot be directly applied for the determination of the exact
value of the dead layer, but may explain the obtained value of $\xi$.

For the analysis of the first minimum position and the dead layer estimation
we may also use the theory developed earlier for the dirty limit, since the
estimated value of $\ell_f\sim\xi_2$. From the analysis of the uniform part
of the solutions obtained in Ref.~[\onlinecite{OurPRB2009}] for dirty SINFS
and SIFNS junctions, it follows that the first minimum of the $J_C(d_f)$
dependence must be at $(d_f-d_{dead})/\xi_2\approx3\pi/4$. If we assume that
the detected minimum of $J_C (d_f)$ is the first one, then, the estimation
yields $d_{dead}^{sat}=d_{max}=1.8\,\mathrm{nm} $. This value is rather
large (it is larger than the oscillations halfperiod), similar to Ref. [%
\onlinecite{WeidesCleanSIFS}]. If we assume that the detected minimum of $%
J_C(d_f)$ is the second one, then the first one must be located at $%
d_f\approx1.4\,\mathrm{nm} $ and the dead layer thickness $%
d_{dead}^{sat}=0.2\,\mathrm{nm} $. There is a small shift of experimental
data for $J_C (d_f)$ at the thickness $d_f\approx1.4\,\mathrm{nm} $, but a
real 0-$\pi$ transition was not detected, possibly, due to the fact that $%
d_{dead}^{sat}<d_{max}$. Generally speaking, if we assume that $d_\mathrm{dead}(d_f)$ jumps from $d_\mathrm{max}$ to $d_\mathrm{dead}^\mathrm{sat}$ at $d_\mathrm{max}$ then one should see a jump on $I_C(d_f)$ dependence. However in our case it not observed, see Fig.\ref{fig:exper}(a) and (b). We suppose that this is related to the fact that $\xi_N\approx\xi_1$.

Figure \ref{fig:exper}(c) contains also calculated curves, plotted for the
same model parameters as in Fig. \ref{fig:exper}(b) but for a cleaner
ferromagnet. Usually the purpose of the experimental investigations of SIFS
structures is the design of a $\pi$-Josephson junction with a large $J_C R_N$%
. $R_N$ is determined mainly by the insulating barrier. Our calculations
show that by increasing $l_f$ one can increase $J_C$ by 1-2 orders of
magnitude, see Fig.\ref{fig:exper}(c). It would also be reasonable to delete
extra normal layers to achieve the $\pi$-phase at a smaller $d_f$, and
consequently, to have larger $J_C$; see also Refs. [%
\onlinecite{Buzdin:PiJJ:JETPL2003,Vasenko}].

\section{Conclusion}

\label{Sec:Conclusion}

The SIFS ferromagnetic tunnel Josephson junction has been investigated
within the framework of the quasiclassical Eilenberger equations, that allow
a description of both, the clean and the dirty limits, as well as an
arbitrary scattering. The Eilenberger function $\Phi_+(x,\mu_f)$ may be
approximated by the simple formula (\ref{Fi_appr}) within the entire range
of considered parameters. The decay length $\xi_1$ and the oscillation
period $2\pi\xi_2$ depends not only on the mean free path $\ell_f$ and the
exchange energy $H$ in the ferromagnet, but also on the ferromagnet
thickness $d_f$, and can be nonmonotonic as a function of $\xi_H$ or $\ell_f$%
.

The approximation of $J_C (d_f)$ by an exponential function or its
combinations has some restrictions. The applicability of the Usadel equation
has been established for $\ell_f \lesssim 0.1\xi_H$.

The developed approach has been used to fit experimental data \cite%
{WeidesCleanSIFS} providing a satisfactory fitting of the $J_C (d_f)$
dependence. It allows to explain the values $\xi_1>\xi_2$ and to give some
practical recommendations on how to increase the $J_C R_N$ product.

\begin{acknowledgments}
We gratefully acknowledge M. Weides for providing
experimental data and helpful discussions, as well as A.B. Granovskiy and
N.S. Perov. This work was supported by the Russian Foundation for Basic
Research (Grants 10-02-90014-Bel-a, 10-02-00569-a, 11-02-12065-ofi-m), by the Deutsche
Forschungsgemeinschaft (DFG) via the SFB/TRR 21 and by project Go-1106/03, by the Deutscher
Akademischer Austauschdienst.
\end{acknowledgments}

\appendix
\section{}

Substituting the expression (\ref{fi_plus_0}) into the Eilenberger equation (%
\ref{FplusinF}), multiplying the obtained equations by $\cos \left[ \pi
k(x+d_{f})/d_{f}\right] ,$ and integrating them over $x$ one can find the
relation between coefficients $Q_{m}$ and $B$ (\ref{BCSF_1}), namely
\begin{equation}
Q_{m}=\frac{\left\langle Q_{m}\right\rangle +\left\langle B\frac{\zeta _{f}}{%
d_{f}}\frac{\left( -1\right) ^{m}}{M}\sinh \frac{d_{f}}{\zeta _{f}}%
\right\rangle }{k_{f}M},~  \label{Qm}
\end{equation}%
where
\begin{equation*}
M =\pi ^{2}m^{2}\frac{\mu ^{2}}{q^{2}}+1, \quad \left\langle \frac{1}{M}%
\right\rangle =\frac{q}{\pi m}\arctan \frac{\pi m}{q}\;.
\end{equation*}%
Averaging both sides of (\ref{Qm}) over the angle $\theta$ we get
\begin{equation}
\left\langle Q_{m}\right\rangle =\left\langle B\frac{\zeta _{f}}{d_{f}}\frac{%
\left( -1\right) ^{m}}{M}\sinh \frac{d_{f}}{\zeta _{f}}\right\rangle \frac{%
\left\langle \frac{1}{k_{f}M}\right\rangle }{\left[ 1-\left\langle \frac{1}{%
k_{f}M}\right\rangle \right] }.  \label{QmAv}
\end{equation}%
A substitution of (\ref{QmAv}) into (\ref{Qm}) finally gives the relation
between the coefficients in expression (\ref{fi_plus_0}).
\begin{equation}
Q_{m}=\frac{\left\langle B\frac{\zeta _{f}}{d_{f}}\frac{\left( -1\right) ^{m}%
}{M}\sinh \frac{d_{f}}{\zeta _{f}}\right\rangle }{M}\frac{1}{\left[
k_{f}-\left\langle \frac{1}{M}\right\rangle \right] }\;.  \label{QmF}
\end{equation}%
Expressions (\ref{QmF}) and (\ref{fi_plus_0}) permit to rewrite the solution
of the Eilenberger equations in the closed
% relative to integrated coefficients $B(\mu _{f})$
form
\begin{widetext}
\begin{equation}
\Phi _{+}=\sum_{m=-\infty }^{\infty }\frac{\left\langle B(\mu )\frac{\left(
-1\right) ^{m}}{M(\mu )}\frac{\mu }{q}\sinh \frac{q}{\mu }\right\rangle
_{\mu }}{M(\mu _{f})\left[ k_{f}-\left\langle \frac{1}{M}\right\rangle %
\right] }\cos \frac{\pi m(x+d_{f})}{d_{f}}+B(\mu _{f})\cosh \left( q\frac{%
x+d_{f}}{d_{f}\mu _{f}}\right)\;,  \label{fi_plus_01}
\end{equation}%
and for an isotropic component $\left\langle \Phi _{+}\right\rangle $ of this solution to get
\begin{equation}
\left\langle \Phi _{+}\right\rangle =k_{f}\sum_{m=-\infty }^{\infty }\frac{%
\left\langle B(\mu )\frac{\left( -1\right) ^{m}}{M(\mu )}\frac{\mu }{q}\sinh
\frac{q}{\mu }\right\rangle _{\mu }}{\left[ k_{f}-\left\langle \frac{1}{M}%
\right\rangle \right] }\cos \frac{\pi m(x+d_{f})}{d_{f}}.  \label{fi_av}
\end{equation}

The form of the solution of the Eilenberger equations in the S layer essentially depends on transport properties of the FS interface.

If the transparency $D(\mu_f)$ is small, then in the first approximation on $D(\mu_f)$, we can neglect the suppression of superconductivity in the S region and consider the order parameter and the Eilenberger functions as constants independent on space coordinates, which are equal to their bulk values, thus leading to the boundary condition for determination of integration constants
$B(\mu _{f})$ in the form of Eq. (\ref{BCSF_1}). From the boundary condition (\ref{BCSF_1}) we get
\begin{equation}
B(\mu _{f})=D(\mu _{f})\frac{\Delta _{0}}{\sqrt{\Delta _{0}^{2}+\omega ^{2}}}%
\frac{1}{\sinh \left( \frac{q}{\mu _{f}}\right) },
\end{equation}
and obtain the analytical solution in the form (\ref{fi_plus}) that, from our point of view, is more convenient for the further analysis than solutions previously used in Refs.~[\onlinecite{Volkov,Linder}].

For the described three forms (\ref{D1})--(\ref{D3}) of $D(\mu _{f})$ dependencies we get
\begin{eqnarray}
\left\langle D(\mu )\frac{\mu }{M(\mu )}\right\rangle _{\mu
} &=& D_{1}\int_{0}^{1}\frac{\mu d\mu }{\pi ^{2}m^{2}\mu ^{2}/q^{2}+1}=\frac{%
D_{1}}{2}\frac{q^{2}}{\pi ^{2}m^{2}}\ln \left( \frac{\pi ^{2}m^{2}}{q^{2}}%
+1\right) ,  \label{D1_av} \\
%\end{equation}%
%\begin{equation}
\left\langle D(\mu )\frac{\mu }{M(\mu )}\right\rangle _{\mu
} &=& D_{2}\int_{0}^{1}\frac{\mu ^{2}d\mu }{\pi ^{2}m^{2}\mu ^{2}/q^{2}+1}=D_{2}%
\frac{q^{2}}{\pi ^{2}m^{2}}\left( 1-\frac{q}{\pi m}\arctan \frac{\pi m}{q}%
\right) ,  \label{D2_av} \\
%\end{equation}%
%\begin{equation}
\left\langle D(\mu )\frac{\mu }{M(\mu )}\right\rangle _{\mu
} &=& D_{3}\int_{0}^{1}\frac{\mu ^{3}d\mu }{\pi ^{2}m^{2}\mu ^{2}/q^{2}+1}=\frac{%
D_{3}}{2}\frac{q^{2}}{\pi ^{2}m^{2}}\left[ 1-\frac{q^{2}}{\pi ^{2}m^{2}}\ln
\left( \frac{\pi ^{2}m^{2}}{q^{2}}+1\right) \right]\;.  \label{D3_av} \\
\end{eqnarray}%
Substituting the averages (\ref{D1_av})--(\ref{D3_av}) into the expression (\ref{fi_plus}) we get the Eilenberger function for different FS interface transparencies (\ref{Fi1})--(\ref{Fi3}). The functions (\ref{Fi1})--(\ref{Fi3}) averaged over the angle have the following forms.

First, for $D(\mu
_{f})=D_{1}$
\begin{equation}
\left\langle \Phi _{+}\right\rangle =\frac{\Delta _{0}D_{1}}{\sqrt{\Delta
_{0}^{2}+\omega ^{2}}}\left[ \frac{1}{2q\left( k_{f}-1\right) }%
+\sum_{m=1}^{\infty }\frac{\frac{q}{2\pi ^{2}m^{2}}\ln \left( \frac{\pi
^{2}m^{2}}{q^{2}}+1\right) }{k_{f}-\frac{q}{\pi m}\arctan \frac{\pi m}{q}}%
\cos \frac{\pi mx}{d_{f}}\right] .  \label{Fi1_av}
\end{equation}%
Second, for $D(\mu _{f})=\mu D_{2}$
\begin{equation}
\left\langle \Phi _{+}\right\rangle =\frac{\Delta _{0}D_{2}}{\sqrt{\Delta
_{0}^{2}+\omega ^{2}}}\left[ \frac{1}{3q\left( k_{f}-1\right) }%
+2\sum_{m=1}^{\infty }\frac{\frac{q}{\pi ^{2}m^{2}}\left( 1-\frac{q}{\pi m}%
\arctan \frac{\pi m}{q}\right) }{k_{f}-\frac{q}{\pi m}\arctan \frac{\pi m}{q}%
}\cos \frac{\pi mx}{d_{f}}\right] .  \label{Fi2_av}
\end{equation}%
Third, for $D(\mu _{f})=\mu ^{2}D_{3}$
\begin{equation}
\left\langle \Phi _{+}\right\rangle =\frac{\Delta _{0}D_{3}}{\sqrt{\Delta
_{0}^{2}+\omega ^{2}}}\left[ \frac{1}{4q\left( k_{f}-1\right) }%
+\sum_{m=1}^{\infty }\frac{\frac{q}{\pi ^{2}m^{2}}\left[ 1-\frac{q^{2}}{\pi
^{2}m^{2}}\ln \left( \frac{\pi ^{2}m^{2}}{q^{2}}+1\right) \right] }{k_{f}-%
\frac{q}{\pi m}\arctan \frac{\pi m}{q}}\cos \frac{\pi mx}{d_{f}}\right] .
\label{Fi3_av}
\end{equation}
\end{widetext}

Our calculation yields the following value of the suppression parameter for
the Usadel equation: $\gamma _{B}=\ell _{f} \left\langle \mu D(\mu
)\right\rangle ^{-1}/ 3\sqrt{d_{f}/2\pi T_{c}} .$ %\begin{equation}
%\gamma _{B}=\frac{1}{3}\frac{\ell _{f}}{\sqrt{\frac{d_{f}}{2\pi T_{c}}}}%
%\left\langle \mu D(\mu )\right\rangle ^{-1} ,
%\end{equation}
It differs from the one obtained early \cite{Kupriyanov88} by a factor 2. We
suppose, that this is a consequence of an approximation we have made in
boundary conditions (\ref{BCSF_1}) at the FS interface. We have neglected
all spatial variations in the S part as well as the back influence of the
ferromagnet on the superconductor, and this factor 2 is the price for this
approximation. If we use the boundary condition (\ref{BCSF_1}) in its full
form \cite{Zaitsev}, we get for $\gamma _{B}$:
\begin{equation*}
\gamma _{B}=\frac{2}{3}\frac{\ell _{f}}{\sqrt{\frac{d_{f}}{2\pi T_{c}}}}%
\left\langle \mu D(\mu )\right\rangle ^{-1}.
\end{equation*}

Our three models of the FS interface yield
\begin{eqnarray}
\gamma _{B}&=&\frac{2}{3}\frac{\ell _{f}}{D_{1}\sqrt{\frac{d_{f}}{2\pi T_{c}}%
}}\left\langle \mu \right\rangle ^{-1}=\frac{4}{3}\frac{\ell _{f}}{D_{1}%
\sqrt{\frac{d_{f}}{2\pi T_{c}}}},  \label{gammaB1} \\
\gamma _{B}&=&\frac{2}{3}\frac{\ell _{f}}{D_{2}\sqrt{\frac{d_{f}}{2\pi T_{c}}%
}}\left\langle \mu ^{2}\right\rangle ^{-1}=2\frac{\ell _{f}}{D_{2}\sqrt{%
\frac{d_{f}}{2\pi T_{c}}}},  \label{gammaB2} \\
\gamma _{B}&=&\frac{2}{3}\frac{\ell _{f}}{D_{3}\sqrt{\frac{d_{f}}{2\pi T_{c}}%
}}\left\langle \mu ^{3}\right\rangle ^{-1}=\frac{8}{3}\frac{\ell _{f}}{D_{3}%
\sqrt{\frac{d_{f}}{2\pi T_{c}}}}.  \label{gammaB3}
\end{eqnarray}

%
%=======================================================

%\bibliographystyle{plain}
\bibliography{pi,SFS3,CleanSF}
% note: SFS3 modified: Weides:2006:SIFS-HiJcPiJJ --> condmat-Reference deleted, as full Ref. is already available; VolkovCritKupr --> added

\end{document}